\documentclass[twocolumn,showpacs,preprintnumbers,amsmath,amssymb,pra]{revtex4}

\usepackage{color}
\usepackage{graphicx}
\usepackage{dcolumn}
\usepackage{bm}
\usepackage[amssymb]{SIunits}

\begin{document}

\title{Geometric phase gates based on stimulated Raman adiabatic passage in tripod systems}
\author{Ditte M\o ller}
\email{dittem@phys.au.dk}
\author{Lars Bojer Madsen}
\author{Klaus M\o lmer}
\affiliation{Lundbeck Foundation Theoretical Center for Quantum System Research,\\
Department of Physics and Astronomy, University of Aarhus, DK-8000,
Denmark.}
\date{\today }

\begin{abstract}
We consider stimulated Raman adiabatic passage (STIRAP) processes in
tripod systems and show how to generate purely geometric phase
changes of the quantum states involved. The geometric phases are
controlled by three laser fields where pulse shapes, relative field
strength and phases can be controlled. We present a robust set of
universal gates for quantum computing based on these geometric
phases: a one-qubit phase gate, a Hadamard gate and a two-qubit
phase gate.
\end{abstract}


\maketitle

\section{Introduction}
With the growing interest in quantum computation and information the
search for efficient and robust quantum gates has become
increasingly important. Deutsch presented in 1989 a three qubit
quantum gate and showed that this gate together with arbitrary
one-qubit rotations are sufficient to create any quantum network
\cite{deutsch89}. Such a set of gates are called universal for
quantum computation. Since then many sets of gates were proven to be
universal \cite{barenco95} - one of them consisting of a one-qubit
phase gate ($S$), a one-qubit Hadamard gate ($H$), and a two-qubit
controlled phase gate ($CS$). Explicitly in the one- and two-qubit bases ($\{\vert 0 \rangle, \vert 1 \rangle\}$, $\{ \vert 0 0 \rangle, \vert 0 1 \rangle, \vert 10 \rangle, \vert 11 \rangle \}$) the form of these gates are
\begin{align}\label{eq:gates}
&S=\begin{bmatrix} 1 & 0 \\ 0 & e^{i\phi_1}
\end{bmatrix}, \quad H=\frac{1}{\sqrt{2}}\begin{bmatrix}1 & 1
\\1 & -1\end{bmatrix},\\\notag
&CS=\frac{1}{\sqrt{2}}\begin{bmatrix}1 & 0 & 0 & 0 \\
0 & 1 & 0 & 0 \\
0 & 0 & 1 & 0 \\
0 & 0 & 0 & e^{i\phi_2}
\end{bmatrix}.
\end{align}
The purpose of the present work is to show that all these gates can
be implemented using only gates based on adiabatic evolution using
stimulated Raman adiabatic passage (STIRAP) and leading to geometric
phases. A quantum system which starts out in the $n$'th eigenstate
of a Hamiltonian that changes adiabatically in time will according
to the adiabatic theorem \cite{messiah61} remain in the $n$'th
eigenstate but may acquire a phase. In 1984 Berry showed that
besides a dynamical part $\theta_n=- \int \omega_n(t) dt$ associated
with the eigenfrequency $\omega_n = E_n / \hbar$ the state also
acquires a geometric phase, $\gamma_n$ \cite{berry84}
\begin{equation}
\Psi(0)=\psi_n(0)\rightarrow\Psi(t)=\exp[i({\theta_n+\gamma_n})]\psi_n(t).
\end{equation}
The geometric part depends on the geometric properties of the
parameter space of the Hamiltonian and Berry showed how to calculate
$\gamma_n$ when the eigenstate is known and the system is
non-degenerate \cite{berry84}. Berry's phase was generalized to
degenerate systems by Wilczek and Zee \cite{wilczek84} and to
non-adiabatic evolution of the system by Aharonov and Anandan
\cite{aharonov87}. Quantum computation relying on these geometric
quantum phases is called holonomic quantum computation
\cite{zanardi99} and is expected to be particularly robust against
noise. Previously, proposals for holonomic quantum computation were
presented for nuclear magnetic resonance \cite{Jones00}, neutral
atoms \cite{recati02} and trapped ions \cite{duan01}.

Control of the geometric phases requires an adiabatic evolution and
preferably an eigenstate with a zero-valued eigenenergy in order to
avoid the build-up of an additional dynamic phase as the system
evolves. A well-described adiabatic process is STIRAP where
population is transferred from one quantum state to another in a
three-level lambda-system when subject to two different laser pulses
ordered in a counterintuitive time-sequence \cite{bergmann98}. The
STIRAP process was shown to be very efficient and robust
theoretically as well as experimentally \cite{oreg84, bergmann98,
gaubatz90, broers92, goldner94, lawall94, cubel05, sorensen06}.
Geometric phases accumulated during a STIRAP process were previously
investigated for tripod systems \cite{unanyan99} and used for
single-qubit rotations \cite{kis02}, entanglement between atoms in a
cavity \cite{pachos04} and holonomic quantum computation with
trapped ions \cite{duan01}. In \cite{dasgupta06} these phases were
considered for an open quantum system. In this work, we present an
experimental implementable set of universal gates based on geometric
phases arising from population transfer in tripod systems and show
explicitly how they depend on the experimental parameters.

The paper is organized as follows: Sec. \ref{sec:tripodsystem}
presents the atomic system under consideration. In Sec.
\ref{sec:onequbitphase} we present a one-qubit phase gate and in
Sec. \ref{sec:Hadamard} the Hadamard gate. The gates are
investigated analytically as well as numerically. In Sec.
\ref{sec:twoqubitphase} we introduce a coupling between the two
qubits and present a two-qubit controlled phase gate.  In
Sec.~\ref{sec:conclusion} we discuss the robustness and conclude.

\section{Tripod system}\label{sec:tripodsystem}
\begin{figure}[htbp]
  \centering
  \includegraphics[width=0.5\textwidth]{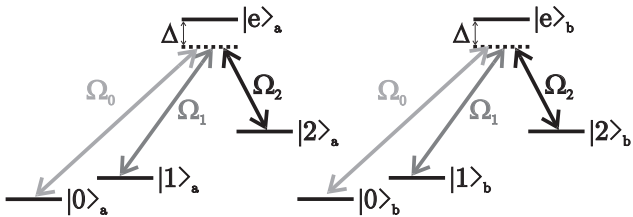}
  \caption{Two atomic four-level tripod systems with three laser fields applied with Rabi frequencies $\Omega_0$, $\Omega_1$, $\Omega_2$. The one-photon detuning is denoted by $\Delta$, and the subscripts $a$ and $b$  refer to two different atoms.}
  \label{fig:tripodsystems}
\end{figure}

We consider two atoms (ions or neutrals) with a tripod level
structure as shown in Fig.~\ref{fig:tripodsystems}. The three lower
states ($|0\rangle$,$|1\rangle$ and $|2\rangle$) are long-lived and
coupled to the upper state $|e\rangle$ by application of three laser
fields with Rabi frequencies $\Omega_0$, $\Omega_1$, $\Omega_2$,
respectively. In practice the lower states can be ground Zeeman- or
hyperfine-sublevels, and $|e\rangle$ is an electronically excited
state or excited state manifold. We assume the laser fields are on
two-photon resonance and denote the one-photon detuning by $\Delta$.
We use STIRAP processes to transfer population among the three lower
states and the excited state $|e\rangle$ is therefore never
populated during the process which ensures that no loss occurs due
to spontaneous emission. We consider $\{|0\rangle,|1\rangle\}$ as
our qubit-states.
In the following we implement the universal set of quantum gates
(\ref{eq:gates}) using geometric phases acquired by transferring
population adiabatically with STIRAP-processes.

\section{One-qubit phase gate} \label{sec:onequbitphase}
\begin{figure}[htbp]
  \centering
  \includegraphics[width=0.5\textwidth]{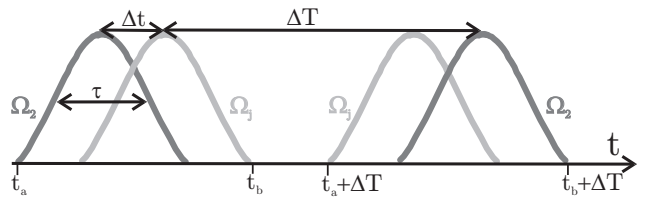}
  \caption{Pulse sequence transferring population from $|j\rangle$
  to $|2\rangle$ and back to $|j\rangle$ again. The FWHM of each of the four pulses is $\tau$ (see Eq.~(\ref{eq:sinpulses})), $\Delta T$ is the delay between the two sequences and $\Delta t$ the delay between two pulses within one sequence.}
  \label{fig:pulses}
\end{figure}

We first consider how to use STIRAP to perform a simple one-qubit
phase gate: $|j\rangle \rightarrow e^{i\phi_j}|j\rangle$ where
$|j\rangle$ is either of the two qubit states $|0\rangle$ and
$|1\rangle$. For this purpose we use a single STIRAP process to
transfer the population from $|j\rangle$ to $|2\rangle$ and another
to transfer the population back to $|j\rangle$ again. During this
process $|j\rangle$ gains a geometric phase. The total pulse
sequence is shown in Fig.~\ref{fig:pulses}.
To solve the time evolution during these two STIRAP sequences we
consider the Hamiltonian for the
\{$|j\rangle$,$|e\rangle$,$|2\rangle$\} lambda system in the
rotating wave approximation (RWA), when we assume two-photon
resonance between $\vert j \rangle$ and $\vert 2 \rangle $ (see
Fig.~\ref{fig:tripodsystems})
\begin{align}\label{eq:hamilton}
H(t)=\frac{\hbar}{2}\left[\begin{array}{ccc}
0 & \Omega_{j}(t) & 0\\
\Omega^{\ast}_{j}(t) & 2\Delta &\Omega_{2}(t) \\
0 & \Omega_{2}^{\ast}(t) & 0
\end{array}\right].
\end{align}
We parameterize the complex Rabi frequencies as
\begin{align}
\Omega_{j}(t)&=\sin\theta(t)\sqrt{|\Omega_{j}(t)|^{2}+|\Omega_{2}(t)|^{2}},\label{eq:omegaj}\\
\Omega_{2}(t)&=\cos\theta(t)\sqrt{|\Omega_{j}(t)|^{2}+|\Omega_{2}(t)|^{2}}e^{i\varphi(t)},\label{eq:omega2}
\end{align}
and diagonalize (\ref{eq:hamilton}) to obtain the energy eigenvalues
\begin{equation}
\omega^{\pm}=\Delta\pm\sqrt{\Delta^{2}+\Omega_{j}^{2}+\Omega_{2}^{2}}
\hspace{0,2cm}, \hspace{1cm}\omega^{D}=0,
\end{equation}
with the non-absorbing zero-valued dark state ($\omega^D=0$) given by
\begin{equation}
|D(t)\rangle=\cos\theta(t)|j\rangle-\sin\theta(t)e^{i\varphi(t)}|2\rangle.\label{eq:dark}
\end{equation}
Now we assume that this state is the initial state
$|\psi(-\infty)\rangle=|D(-\infty)\rangle$ at time $t=-\infty$
before the pulses and that we vary the real amplitudes of
$\Omega_j(t)$ and $\Omega_2(t)$ and the phase $\varphi(t)$ of
$\Omega_2(t)$ in an adiabatic way such that all population stays in
$|D(t)\rangle$. During this evolution the $|D(t)\rangle$-state will
pick up a phase which is purely geometric because the state has zero
energy eigenvalue. The phase will depend on the evolution of the
parameters $\theta$ and $\varphi$ and we define these as a vector,
$\vec{R}=(\theta(t),\varphi(t))$, in parameter-space of the
Hamiltonian. Then the acquired Berry phase is exactly the integral
\begin{equation}
\gamma_{n_1}=i\int_{\bar{R}_i}^{\bar{R}_f}\langle D|
\nabla_{\bar{R}}|D\rangle \cdot
d\bar{R}=-\int_{\varphi(t_i)}^{\varphi(t_f)} \sin^2\theta(t)
d\varphi(t),\label{eq:gamma1}
\end{equation}
where $\sin^2\theta$ is found from Eq. (\ref{eq:omegaj})
\begin{equation}
\sin^2\theta(t)=\frac{\Omega^2_j(t)}{|\Omega_j(t)|^2+|\Omega_2(t)|^2}.\label{eq:sin2}
\end{equation}
In the pulse sequence of Fig.~\ref{fig:pulses} we assume that all
four pulses are described by the common function $\Omega(t)$. In
addition to $\Omega(t)$, the Rabi frequency $\Omega_2(t)$ is defined
by the phase $\varphi(t)$ (see Eq. (\ref{eq:omega2})). The instants
of time $t_a$ and $t_b$ in Fig. \ref{fig:pulses} are defined such
that $\sin^2\theta(t)\approx 0$ for $t<t_a$ $\vee$ $t>t_b+\Delta T$
and $\sin^2\theta(t)\approx 1$ for $t_b<t<t_a+\Delta T$. With these
definitions we obtain from Eq. (\ref{eq:gamma1}) and (\ref{eq:sin2})
\begin{align}
\gamma_{n_1}=&-\int_{\varphi(t_a)}^{\varphi(t_b)}
\frac{\Omega^2(t)}{\Omega^2(t)+\Omega^2(t+\Delta t)} d\varphi\\
\notag &-\int_{\varphi(t_b)}^{\varphi(t_a+\Delta T)} 1 d\varphi\\
\notag &-\int_{\varphi(t_a+\Delta T)}^{\varphi(t_b+\Delta T)}
\frac{\Omega^2(t+\Delta t-\Delta T)}{\Omega^2(t+\Delta t-\Delta
T)+\Omega^2(t-\Delta T)} d\varphi.
\end{align}
Substituting $t'=t-\Delta T$ in the last integral and assuming that
$\varphi$ is a monotonic function we obtain
\begin{align}\label{eq:gamma1calc}
\gamma_{n_1}=&-\int_{t_a}^{t_b}
\frac{\Omega^2(t)}{\Omega^2(t)+\Omega^2(t+\Delta t)}\cdot \frac{d\varphi}{dt} dt \\
\notag &-\int_{t_b}^{t_a+\Delta T} \frac{d\varphi}{dt} dt \\
\notag &-\int_{t_a}^{t_b} \frac{\Omega^2(t'+\Delta
t)}{\Omega^2(t'+\Delta t)+\Omega^2(t')} \cdot \frac{d\varphi}{dt} dt'\\
\notag =&-\int_{t_a}^{t_a+\Delta
T}\frac{d\varphi}{dt}dt=\varphi(t_a)-\varphi(t_a+\Delta T).
\end{align}
The geometric phase thus only depends on the laser field phases and
requires control of $\Delta T$, the fraction
$\frac{\Omega_j(t)}{\Omega_2(t)}$ and similarity of the four pulses.
All these quantities are routinely controlled to high precision in
present-day laboratories. After the evolution the final state
is\begin{equation}
|\psi_f(t)\rangle=e^{i\gamma_{n_1}}|j\rangle.\label{eq:finalstate}
\end{equation}
\begin{figure}[htbp]
  \centering
  \includegraphics[width=0.5\textwidth]{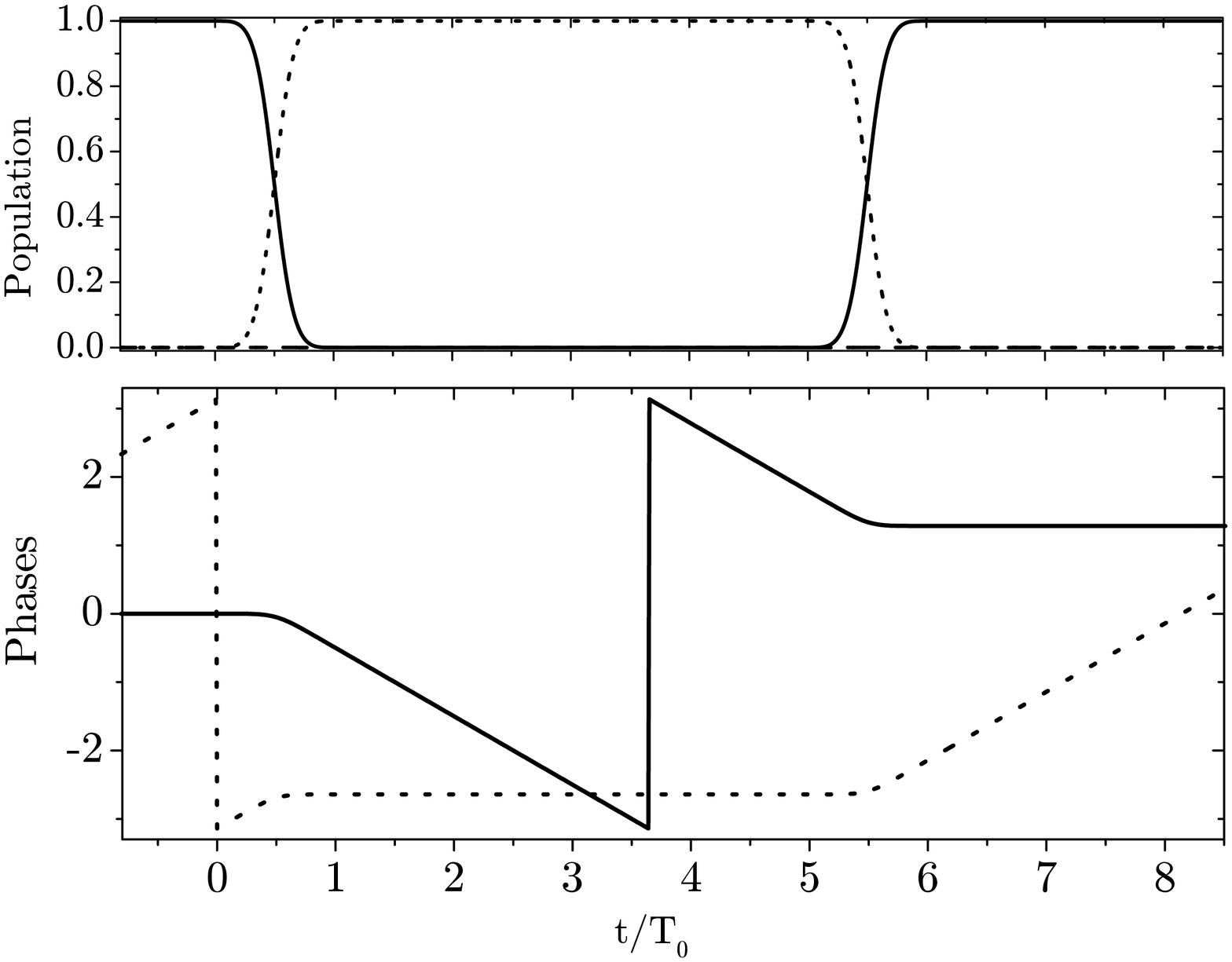}
  \caption{
The upper panel shows the evolution of the population of states
$|j\rangle$ (full), $|e\rangle$ (dashed), and $|2\rangle$ (dotted).
The lower panel shows the evolution of the phases, $\phi_{j}$ of
state $|j\rangle$ (full) and $\phi_{2}$ of state $|2\rangle$
(dotted). Numerical and analytical results cannot be distinguished
on the scale of the figure. The calculations were made with
sin$^2$-pulses (\ref{eq:sinpulses}), $\varphi=t/T_0$ and parameters:
$\Omega_{\textrm{max},j}/2\pi=\Omega_{\textrm{max},2}/2\pi=100/T_0$,
$\Delta t/T_0=1$, $\tau/T_0=1$, $\Delta T/T_0=5$.}
  \label{fig:stirapphaseback}
\end{figure}
The population and the phases of the three states $|j\rangle$,
$|e\rangle$ and $|2\rangle$ can be found numerically by solving the
time-dependent Schr\"{o}dinger equation. In addition the phases can
be calculated analytically as described above (Eq.
(\ref{eq:gamma1calc}) and (\ref{eq:finalstate})). In
Fig.~\ref{fig:stirapphaseback} we show the population of the states
as well as the evolution of their phases when we assume
$\varphi=t/T_0 \Rightarrow \gamma_{n_1}=-\Delta T/T_0$ and use
$\sin^2$-pulses with same amplitude
\begin{equation}
\Omega(t)=\begin{cases}
\Omega_{\textrm{max}}\sin^{2}(\frac{\pi t}{2 \tau}) & \text{if } 0<t<2\tau\\
0 & \text{otherwise}
\end{cases}.\label{eq:sinpulses}
\end{equation}
The factor of $2$ in the pulse assures that $\tau$ corresponds to
the FWHM. The populations are shown in the upper panel of
Fig.~\ref{fig:stirapphaseback}. Initially all population is in the
$|j\rangle$-state (full curve). During the first STIRAP-process the
population is transferred from $|j\rangle$ to $|2\rangle$ (dotted
curve) while the second STIRAP-process transfers the population back
to $|j\rangle$. The electronic excited $|e\rangle$-state (dashed
curve) is never populated and hence no loss of population occurs due
to spontaneous emission from $|e\rangle$. The evolution of the
phases is shown in the lower panel of
Fig.~\ref{fig:stirapphaseback}. The full curve shows the phase of
$|j\rangle$. This phase is the geometric phase
$\phi_{j}=\gamma_{n_1}$. The phase $\phi_{j}$ remains zero until the
first set of STIRAP pulses arrive ($t_a$ in Fig.~\ref{fig:pulses});
it then accumulates a phase until the second pair of pulses has
passed ($t_b+\Delta T$ in Fig.~\ref{fig:pulses}). The total acquired
phase is as shown in Eq. (\ref{eq:gamma1calc}),
$\gamma_{n_1}=\varphi(t_a)-\varphi(t_a+\Delta T)=-\Delta T/T_0$. The
phase of the $|2\rangle$-state (dotted curve) contains not only the
geometric phase $\gamma_{n_1}$ but also the additional
$\varphi(t)-\pi$ (See Eq. \ref{eq:dark}) yielding a total phase
$\phi_{2}=\gamma_{n_1}+\varphi(t)-\pi$. The $|2\rangle$-state
therefore accumulates a phase before the first and after the second
STIRAP process -but non in between where $\gamma_{n_1}$ and
$\varphi(t)$ cancel each other. In the time-windows where the states
in question are populated the direct numerical solution of the
time-dependent Schr\"{o}dinger equation gives results for the phases
in agreement with the above analytical results.

\section{Hadamard gate}\label{sec:Hadamard}
To implement a Hadamard gate we use three laser fields with Rabi
frequencies $\Omega_0$, $\Omega_1$ and $\Omega_2$ (See
Fig.~\ref{fig:tripodsystems}). We apply $\Omega_0$ and $\Omega_1$ in
a constant ratio $\tan\theta_{01}=\frac{|\Omega_0|}{|\Omega_1|}$
with phase difference $\phi_{01}$ and assume two-photon resonance.
In this case the dressed states of the system are a single dark
($|D_H \rangle$) and two bright states ($|+ \rangle$,$|-\rangle$)
\begin{align}\label{eq:dressedstates}
|D_H \rangle=&\cos\theta_{01}(t)|0\rangle-\sin\theta_{01}(t)e^{i\phi_{01}}|1\rangle,\\
\notag
|+ \rangle=&\frac{1}{\sqrt{2}}\left(\sin\delta(\sin\theta_{01}(t)|0\rangle+\cos\theta_{01}(t)e^{i\phi_{01}}|1\rangle\right.)\\
\notag
&\left.+\cos\delta|e\rangle\right),\\
\notag |-\rangle=&\frac{1}{\sqrt{2}}\left(\cos\delta(\sin\theta_{01}(t)|0\rangle+\cos\theta_{01}(t)e^{i\phi_{01}}|1\rangle\right.)\\
\notag &\left.-\sin\delta|e\rangle\right),\\\notag
\end{align}
with eigenvalues $\omega_{D_H}=0$,
$\omega_{\pm}=\Delta\pm\sqrt{\Delta^2+\Omega_0^2+\Omega_1^2}$ and
$\delta$ defined by $\tan\delta=\sqrt{\frac{-\omega_-}{\omega_+}}$.
To ensure that no population is transferred to the $|e\rangle$-state
we choose as our initial state
\begin{align}\label{eq:instate}
|\psi_i\rangle=&a|D_{H}\rangle+\sqrt{2}b(\sin\delta|+\rangle+\cos\delta|- \rangle)\\
\notag
=&a|D_{H}\rangle+b(\sin\theta_{01}|0\rangle+\cos\theta_{01}e^{i\phi_{01}}|1\rangle)\\\notag
=&a|D_{H}\rangle+b|B\rangle,\\\notag
\end{align}
where $a,b$ are normalization constants and
$|B\rangle=\sin\theta_{01}|0\rangle+\cos\theta_{01}e^{i\phi_{01}}|1\rangle$
is the bright part of $|\psi_i\rangle$. We now use the same pulse
sequence as in the case of the one-qubit phase gate
(Fig.~\ref{fig:pulses}) but with both $\Omega_0$ and $\Omega_1$
applied. The dark state $|D_{H}\rangle$ does not couple to
$|2\rangle$ and is therefore unaffected by the pulses, while the
bright state $|B\rangle$ does couple and is therefore transferred to
$|2\rangle$ and back again acquiring a geometric phase. In this
sense the dynamics is similar to the one-qubit case and accordingly
$|B\rangle\rightarrow e^{i\gamma_{n_H}}|B\rangle$ with
$\gamma_{n_H}=-\int\sin^2\theta_H d\varphi_H$. The geometric phase
$\gamma_{n_H}$ is controlled by
$\tan\theta_H=\frac{\sqrt{|\Omega_0|^2+|\Omega_1|^2}}{|\Omega_2|}$
and $\varphi_H$, which is the phase difference between $\Omega_0$
and $\Omega_2$. After the pulses the system then ends up in
\begin{align}\label{eq:fistate}
|\psi_f\rangle&=a|D_H\rangle+be^{i\gamma_{n_H}}|B\rangle\\\notag
&=a|D_H\rangle+be^{i\gamma_{n_H}}(\sin\theta_{01}|0\rangle+\cos\theta_{01}e^{i\phi_{01}}|1\rangle).\\\notag
\end{align}
To implement the Hadamard gate we control the Rabi frequencies to
obtain
$\Omega_{\textrm{max},2}=\sqrt{\Omega_{\textrm{max},0}^2+\Omega_{\textrm{max},1}^2}$
and $\Omega_0=-(\sqrt{2}-1)\Omega_1$. This latter relation leads to
$\theta_{01}=\frac{\pi}{8}$ and $\phi_{01}=\pi$. We control the
pulse sequence such that $\gamma_{n_H}=-\pi$ and with these
parameters we find from (\ref{eq:dressedstates}), (\ref{eq:instate})
and (\ref{eq:fistate})
\begin{align}
|\psi_i\rangle=&(a\cos{\mbox{$\frac{\pi}{8}$}}+b\sin{\mbox{$\frac{\pi}{8}$}})|0\rangle+(a\sin{\mbox{$\frac{\pi}{8}$}}-b\cos{\mbox{$\frac{\pi}{8}$}})|1\rangle\\\notag
|\psi_f\rangle=&(a\cos{\mbox{$\frac{\pi}{8}$}}-b\sin{\mbox{$\frac{\pi}{8}$}})|0\rangle+(a\sin{\mbox{$\frac{\pi}{8}$}}+b\cos{\mbox{$\frac{\pi}{8}$}})|1\rangle\\\notag
\end{align}
Now the initial condition $|\psi_i\rangle=|0\rangle$ corresponds to
$(a,b)=(\cos(\frac{\pi}{8}),\sin(\frac{\pi}{8}))$ yielding a final
state $|\psi_f\rangle=\frac{1}{\sqrt{2}}(|0\rangle+|1\rangle)$,
while $|\psi_i\rangle=|1\rangle$ corresponds to
$(a,b)=(\sin(\frac{\pi}{8}),-\cos(\frac{\pi}{8}))$ yielding a final
state $|\psi_f\rangle=\frac{1}{\sqrt{2}}(|0\rangle-|1\rangle)$. With
this degree of control we therefore produce a Hadamard gate with
certainty.
 This gate combined with the one-qubit phase gate can generate
arbitrary qubit rotations and the gate is robust as it depends on
controllable parameters such as the pulse shapes, the ratio of Rabi
frequencies, the delay between the two sequences and the phases of
the laser fields.

\section{Two-qubit phase gate}\label{sec:twoqubitphase}
To create a two-qubit phase gate a coupling between the two qubits
is necessary. We consider a coupling $E|22\rangle\langle22|$, where
$E$ is the coupling strength. In the end of this section, we briefly
discuss how such a coupling can be realized. We assume real Rabi
frequencies and all laser fields on resonance. If we wish to solve
the full system of Fig.~\ref{fig:tripodsystems} analytically it is
an advantage to go into the interaction picture with respect to
$H_0=E|22\rangle \langle22|$. In this picture the system has a
six-dimensional null-space yielding six orthonormal dark states,
$|D_i\rangle$ $(i=1,\dots,6)$:
\begin{align}
|D_1\rangle =&|00\rangle,\\\notag |D_2\rangle
=&-\cos\theta_2|10\rangle+\sin\theta_2|20\rangle,\\\notag
|D_3\rangle
=&-\cos\theta_2|01\rangle+\sin\theta_2|02\rangle,\\\notag
|D_4\rangle
=&\frac{1}{\sqrt{2}}(\sin\theta_2(|1e\rangle-|e1\rangle)+\cos\theta_2(|2e\rangle-|e2\rangle)),\\\notag
|D_5\rangle
=&\cos^2\theta_2|11\rangle-\sin\theta_2\cos\theta_2(|12\rangle+|21\rangle)\\
\notag &+\sin^2\theta_2 e^{iEt}|22\rangle,\\\notag |D_6\rangle
=&\frac{1}{\sqrt{2}}(-\sin^2\theta_2|11\rangle-\sin\theta_2\cos\theta_2(|12\rangle+|21\rangle)+|ee\rangle
\\ \notag &-\cos^2\theta_2 e^{iEt}|22\rangle).\\\notag
\end{align}
In this degenerate case we use the method described by Wilczek and
Zee \cite{wilczek84} to find the geometric phases. We assume that we
start with all population in $|11\rangle$ and that only $\Omega_2$
is applied and write $\psi_I(-\infty)=|D_5(-\infty)\rangle$, where
the index I indicates that we are solving the problem in the
interaction picture. When we assume an adiabatic evolution the
population stays within the null-space and $\psi_I(t)$ can be
written as (see also Ref.~\cite{unanyan99})
\begin{equation}
\psi_I(t)=\sum_{b}B_{b}(t)|D_b(t)\rangle.\label{eq:interactionstate}
\end{equation}
The time evolution is given by the time-dependent Schr\"{o}dinger
equation (dot denotes differentiation with respect to time)
\begin{align}\notag
&\dot{\psi}_I(t)=\frac{-i}{\hbar}(H(t)-H_0(t))\psi_I(t)=0 \Rightarrow\\
\notag
&\sum_{b}\dot{B}_{b}(t)D_b(t)\rangle+B_{b}(t)|\dot{D}_b(t)\rangle=0\Rightarrow\\
&\sum_{b}\dot{B}_{b}(t)|D_b(t)\rangle=-\sum_{b}B_{b}(t)|\dot{D}_b(t)\rangle.\\\notag
\end{align}
Here we have used that $(H(t)-H_0(t))|D_b(t)\rangle=0$ for all dark
states. Taking the inner product with $\langle D_c(t)|$ yields
\begin{equation}
\dot{B}_c(t)=-\sum_{b}B_{b}(t)\langle D_c(t)|\dot{D}_b(t)\rangle.
\label{diffligninger}
\end{equation}
The only non-zero $\langle D_c(t)|\dot{D}_b(t)\rangle$-elements are
\begin{align}
\langle D_5(t)|\dot{D}_5(t)\rangle &=iE\sin^4\theta, \\ \notag
\langle D_5(t)|\dot{D}_6(t)\rangle &=\frac{i}{\sqrt{2}}E\cos^2\theta_2\sin^2\theta_2,\\
\notag \langle D_6(t)|\dot{D}_5(t)\rangle
&=\frac{i}{\sqrt{2}}E\cos^2\theta_2\sin^2\theta_2, \\ \notag \langle
D_6(t)|\dot{D}_6(t)\rangle &=\frac{i}{2}E\cos^4\theta_2. \\ \notag
\end{align}
The differential equations for the $B$-coefficients now reduce to
\begin{align}\label{eq:diffb}
\dot{B}_1(t)&=0\textrm{, }\dot{B}_2(t)=0\textrm{,
}\dot{B}_3(t)=0\textrm{, }\dot{B}_4(t)=0,\\ \notag
\dot{B}_5(t)&=-iE\sin^4\theta_2 B_5(t)-\frac{i}{\sqrt{2}}E\cos^2\theta_2\sin^2\theta_2 B_6(t),\\
\notag
\dot{B}_6(t)&=-\frac{i}{\sqrt{2}}E\cos^2\theta_2\sin^2\theta_2
B_5(t)-i\frac{1}{2}E\cos^4\theta_2 B_6(t).\\\notag
\end{align}
Starting with all initial population in $|D_5(t)\rangle$,
Eq.~(\ref{eq:diffb}) shows that $|D_6(t)\rangle$ is populated when
$\cos^2\theta_2\sin^2\theta_2$ is non-vanishing, which is only the
case during the turn-on and turn-off of pulses. Choosing small pulse
widths therefore assures that effectively all population stays in
$|D_5(t)\rangle$ while accumulating the phase. A pulse with FWHM
$\tau=$\unit{1}{\micro\second} results in the transfer of $0.5\%$ of
the population to $|D_6(t)\rangle$ while the smaller FWHM value
$\tau=$\unit{0.5}{\micro\second} reduces this population to $0.1\%$.
The value $\tau=$\unit{0.5}{\micro\second} is experimentally
feasible but requires higher Rabi frequencies. Now, in this regime
where $B_6(t)\approx0$, we may readily solve Eq.~(\ref{eq:diffb})
for $B_5(t)$ and from \ref{eq:interactionstate} we obtain:
\begin{align}
\psi_I(t)=&e^{-iE\int_{-\infty}^{t_f}\sin^4\theta_2 dt}|D_5(t)\rangle\\
\notag =&\cos^2\theta_2 e^{-iE\int_{-\infty}^{t_f}\sin^4\theta_2
dt}|11\rangle\\\notag &-\sin\theta_2\cos\theta_2
e^{-iE\int_{-\infty}^{t_f}\sin^4\theta_2 dt}(|12\rangle+|21\rangle)\\
\notag &+\sin^2\theta_2 e^{-iE\int_{-\infty}^t\sin^4\theta_2
dt+iEt}|22\rangle.\\\notag
\end{align}
Going back to the Schr\"{o}dinger picture an extra phase associated
with the $\vert 22 \rangle$-energy shift is introduced and
\begin{align}
\psi(t)=&e^{-iE|22\rangle\langle22|t}\psi_I(t)\\\notag
=&e^{-iE\int_{-\infty}^t\sin^4\theta_2 dt}\left(\cos^2\theta_2
|11\rangle\right.\\\notag &\left.-\sin\theta_2\cos\theta_2
(|12\rangle+|21\rangle)+\sin^2\theta_2 |22\rangle\right).
\end{align}
With all population initially in $|11\rangle$ and application of the
STIRAP pulse sequence (\ref{fig:pulses}) to both atoms but with only
$\Omega_1$ in Fig.~\ref{fig:tripodsystems} applied, we end up in
$e^{\gamma_{n_2}}|11\rangle$ after the pulse sequence, where
$\gamma_{n_2}=-E\int_{-\infty}^{t_f}\sin^4\theta_2 dt$. If we, in
addition, keep the phases of the laser fields fixed no single-qubit
phases are accumulated, i.e.,
\begin{align}
|00\rangle &\rightarrow |00\rangle\\ \notag |01\rangle &\rightarrow
|01\rangle\\ \notag |10\rangle &\rightarrow |10\rangle\\ \notag
|11\rangle &\rightarrow e^{i\gamma_{n_2}}|11\rangle.\\ \notag
\end{align}

So far, our analysis of the two-qubit gate is general and applicable
to various atomic systems such as optically trapped neutral atoms,
trapped ions and rare-earth ions doped into crystals. To provide the
coupling $E|22\rangle\langle22|$ we need to be a little more
specific, and we suggest in the case of trapped atoms or ions to
exploit the long-range dipole-dipole interaction between Rydberg
excited atoms \cite{Jaksch00} and in doped crystals to exploit the
interaction between excited states with permanent dipole moments
\cite{Ohlsson02}. In \cite{Jaksch00, Ohlsson02} the interactions are
used for quantum gates in stepwise schemes, where first one atom is
excited, and then the interaction blocks the excitation of the
second atom, leading to an entanglement between them. This stepwise
process is not compatible with our adiabatic protocol, and we
suggest instead to apply the interaction to perturb the energy
levels of both atoms when their $|2\rangle$ states are coupled
off-resonantly to the excited states. As shown in
\cite{Bouchoule02}, this off-resonant excitation causes an energy
shifts (AC Stark shifts) of the $|2\rangle$ states, and due to the
dipole-dipole interaction, this shift will have a non-separable
component of precisely the desired form for suitable choices of the
laser detunings and strengths. The energy shift $E$ is given by a
fourth order expansion in Rabi frequencies with a product of three
detunings in the denominator, and the gate thus requires relatively
long interaction times to avoid population transfer to the excited
states.

\section{Conclusions}\label{sec:conclusion}
We have shown that in the adiabatic limit population transfer in
tripod systems introduces purely geometric phases. These phases can
be used in quantum information science to form a set of robust
geometric gates. The performance of the three gates (\ref{eq:gates})
depends on the robustness of the phases:
$\gamma_{n_1}=-\int_{\varphi_i}^{\varphi_f}\sin^2\theta d\varphi$,
$\gamma_{n_H}=-\int_{\varphi_{H_i}}^{\varphi_{H_f}}\sin^2\theta_H
d\varphi_H$ and $\gamma_{n_2}=-E\int_{-\infty}^{t_f}\sin^4\theta_2
dt$. The population transfers are done without ever populating the
upper state $|e\rangle$ in the tripod system (see Fig.
\ref{fig:tripodsystems}), which ensures that the gates are
insensitive to spontaneous emission. Pulse shapes, delay between
sequences and ratio between Rabi frequencies are routinely
controlled experimentally without drift in the laboratory. If we
assume systems where the three laser frequencies lie so close that
all fields can be generated from the same source, the relative
phases of the fields are also easily controllable. This could be
achieved when $\{|0\rangle,|1\rangle,|2\rangle\}$ are atomic Zeeman-
or hyperfine-substates. In this case all gates are very robust with
respect to parameter fluctuations and can be implementable in
present-day laboratories.

\begin{acknowledgments}
This work is supported by the Danish Research Agency (Grant. No.
2117-05-0081).
\end{acknowledgments}


\begin{thebibliography}{22}
\expandafter\ifx\csname natexlab\endcsname\relax\def\natexlab#1{#1}\fi
\expandafter\ifx\csname bibnamefont\endcsname\relax
  \def\bibnamefont#1{#1}\fi
\expandafter\ifx\csname bibfnamefont\endcsname\relax
  \def\bibfnamefont#1{#1}\fi
\expandafter\ifx\csname citenamefont\endcsname\relax
  \def\citenamefont#1{#1}\fi
\expandafter\ifx\csname url\endcsname\relax
  \def\url#1{\texttt{#1}}\fi
\expandafter\ifx\csname urlprefix\endcsname\relax\def\urlprefix{URL }\fi
\providecommand{\bibinfo}[2]{#2}
\providecommand{\eprint}[2][]{\url{#2}}

\bibitem[{\citenamefont{Deutsch}(1989)}]{deutsch89}
\bibinfo{author}{\bibfnamefont{D.}~\bibnamefont{Deutsch}},
  \bibinfo{journal}{Proc. R. Soc. London, Ser. A}
  \textbf{\bibinfo{volume}{425}}, \bibinfo{pages}{73} (\bibinfo{year}{1989}).

\bibitem[{\citenamefont{Barenco et~al.}(1995)\citenamefont{Barenco, Bennett,
  Cleve, Divincenzo, Margolus, Shor, Sleator, Smolin, and
  Weinfurter}}]{barenco95}
\bibinfo{author}{\bibfnamefont{A.}~\bibnamefont{Barenco}},
  \bibinfo{author}{\bibfnamefont{C.~H.} \bibnamefont{Bennett}},
  \bibinfo{author}{\bibfnamefont{R.}~\bibnamefont{Cleve}},
  \bibinfo{author}{\bibfnamefont{D.~P.} \bibnamefont{Divincenzo}},
  \bibinfo{author}{\bibfnamefont{N.}~\bibnamefont{Margolus}},
  \bibinfo{author}{\bibfnamefont{P.}~\bibnamefont{Shor}},
  \bibinfo{author}{\bibfnamefont{T.}~\bibnamefont{Sleator}},
  \bibinfo{author}{\bibfnamefont{J.~A.} \bibnamefont{Smolin}},
  \bibnamefont{and}
  \bibinfo{author}{\bibfnamefont{H.}~\bibnamefont{Weinfurter}},
  \bibinfo{journal}{Phys. Rev. A} \textbf{\bibinfo{volume}{52}},
  \bibinfo{pages}{3457} (\bibinfo{year}{1995}).

\bibitem[{\citenamefont{Messiah}(1961)}]{messiah61}
\bibinfo{author}{\bibfnamefont{A.}~\bibnamefont{Messiah}},
  \emph{\bibinfo{title}{Quantum Mechanics}}, vol.~\bibinfo{volume}{2}
  (\bibinfo{publisher}{North-Holland publishing company},
  \bibinfo{year}{1961}).

\bibitem[{\citenamefont{Berry}(1984)}]{berry84}
\bibinfo{author}{\bibfnamefont{M.~V.} \bibnamefont{Berry}},
  \bibinfo{journal}{Proc. R. Soc. London, Ser. A}
  \textbf{\bibinfo{volume}{392}}, \bibinfo{pages}{45} (\bibinfo{year}{1984}).

\bibitem[{\citenamefont{Wilczek and Zee}(1984)}]{wilczek84}
\bibinfo{author}{\bibfnamefont{F.}~\bibnamefont{Wilczek}} \bibnamefont{and}
  \bibinfo{author}{\bibfnamefont{A.}~\bibnamefont{Zee}},
  \bibinfo{journal}{Phys. Rev. Lett.} \textbf{\bibinfo{volume}{52}},
  \bibinfo{pages}{2111} (\bibinfo{year}{1984}).

\bibitem[{\citenamefont{Aharonov and Anandan}(1987)}]{aharonov87}
\bibinfo{author}{\bibfnamefont{Y.}~\bibnamefont{Aharonov}} \bibnamefont{and}
  \bibinfo{author}{\bibfnamefont{J.}~\bibnamefont{Anandan}},
  \bibinfo{journal}{Phys. Rev. Lett.} \textbf{\bibinfo{volume}{58}},
  \bibinfo{pages}{1593} (\bibinfo{year}{1987}).

\bibitem[{\citenamefont{Zanardi and Rasetti}(1999)}]{zanardi99}
\bibinfo{author}{\bibfnamefont{P.}~\bibnamefont{Zanardi}} \bibnamefont{and}
  \bibinfo{author}{\bibfnamefont{M.}~\bibnamefont{Rasetti}},
  \bibinfo{journal}{Phys. Lett. A} \textbf{\bibinfo{volume}{264}},
  \bibinfo{pages}{94} (\bibinfo{year}{1999}).

\bibitem{Jones00}
J. A. Jones, V. Vedral, A. Ekert, G. Castagnoli, Nature
\textbf{403}, 869 (2000).

\bibitem[{\citenamefont{Recati et~al.}(2002)\citenamefont{Recati, Calarco,
  Zanardi, Cirac, and Zoller}}]{recati02}
\bibinfo{author}{\bibfnamefont{A.}~\bibnamefont{Recati}},
  \bibinfo{author}{\bibfnamefont{T.}~\bibnamefont{Calarco}},
  \bibinfo{author}{\bibfnamefont{P.}~\bibnamefont{Zanardi}},
  \bibinfo{author}{\bibfnamefont{J.~I.} \bibnamefont{Cirac}}, \bibnamefont{and}
  \bibinfo{author}{\bibfnamefont{P.}~\bibnamefont{Zoller}},
  \bibinfo{journal}{Phys. Rev. A} \textbf{\bibinfo{volume}{66}},
  \bibinfo{pages}{032309} (\bibinfo{year}{2002}).

\bibitem[{\citenamefont{Duan et~al.}(2001)\citenamefont{Duan, Cirac, and
  Zoller}}]{duan01}
\bibinfo{author}{\bibfnamefont{L.~M.} \bibnamefont{Duan}},
  \bibinfo{author}{\bibfnamefont{J.~I.} \bibnamefont{Cirac}}, \bibnamefont{and}
  \bibinfo{author}{\bibfnamefont{P.}~\bibnamefont{Zoller}},
  \bibinfo{journal}{Science} \textbf{\bibinfo{volume}{292}},
  \bibinfo{pages}{1695} (\bibinfo{year}{2001}).

\bibitem[{\citenamefont{Bergmann et~al.}(1998)\citenamefont{Bergmann, Theuer,
  and Shore}}]{bergmann98}
\bibinfo{author}{\bibfnamefont{K.}~\bibnamefont{Bergmann}},
  \bibinfo{author}{\bibfnamefont{H.}~\bibnamefont{Theuer}}, \bibnamefont{and}
  \bibinfo{author}{\bibfnamefont{B.~W.} \bibnamefont{Shore}},
  \bibinfo{journal}{Rev. Mod. Phys.} \textbf{\bibinfo{volume}{70}},
  \bibinfo{pages}{1003} (\bibinfo{year}{1998}).

\bibitem[{\citenamefont{Oreg et~al.}(1984)\citenamefont{Oreg, Hioe, and
  Eberly}}]{oreg84}
\bibinfo{author}{\bibfnamefont{J.}~\bibnamefont{Oreg}},
  \bibinfo{author}{\bibfnamefont{F.~T.} \bibnamefont{Hioe}}, \bibnamefont{and}
  \bibinfo{author}{\bibfnamefont{J.~H.} \bibnamefont{Eberly}},
  \bibinfo{journal}{Phys. Rev. A} \textbf{\bibinfo{volume}{29}},
  \bibinfo{pages}{690} (\bibinfo{year}{1984}).

\bibitem[{\citenamefont{Gaubatz et~al.}(1990)\citenamefont{Gaubatz, Rudecki,
  Schiemann, and Bergmann}}]{gaubatz90}
\bibinfo{author}{\bibfnamefont{U.}~\bibnamefont{Gaubatz}},
  \bibinfo{author}{\bibfnamefont{P.}~\bibnamefont{Rudecki}},
  \bibinfo{author}{\bibfnamefont{S.}~\bibnamefont{Schiemann}},
  \bibnamefont{and} \bibinfo{author}{\bibfnamefont{K.}~\bibnamefont{Bergmann}},
  \bibinfo{journal}{J. Chem. Phys.} \textbf{\bibinfo{volume}{92}},
  \bibinfo{pages}{5363} (\bibinfo{year}{1990}).

\bibitem[{\citenamefont{Broers et~al.}(1992)\citenamefont{Broers, van Linden
  van~den Heuvell, and Noordam}}]{broers92}
\bibinfo{author}{\bibfnamefont{B.}~\bibnamefont{Broers}},
  \bibinfo{author}{\bibfnamefont{H.~B.} \bibnamefont{van Linden van~den
  Heuvell}}, \bibnamefont{and} \bibinfo{author}{\bibfnamefont{L.~D.}
  \bibnamefont{Noordam}}, \bibinfo{journal}{Phys. Rev. Lett.}
  \textbf{\bibinfo{volume}{69}}, \bibinfo{pages}{2062} (\bibinfo{year}{1992}).

\bibitem[{\citenamefont{Goldner et~al.}(1994)\citenamefont{Goldner, Gerz,
  Spreeuw, Rolston, Westbrook, Phillips, Marte, and Zoller}}]{goldner94}
\bibinfo{author}{\bibfnamefont{L.~S.} \bibnamefont{Goldner}},
  \bibinfo{author}{\bibfnamefont{C.}~\bibnamefont{Gerz}},
  \bibinfo{author}{\bibfnamefont{R.~J.~C.} \bibnamefont{Spreeuw}},
  \bibinfo{author}{\bibfnamefont{S.~L.} \bibnamefont{Rolston}},
  \bibinfo{author}{\bibfnamefont{C.~I.} \bibnamefont{Westbrook}},
  \bibinfo{author}{\bibfnamefont{W.~D.} \bibnamefont{Phillips}},
  \bibinfo{author}{\bibfnamefont{P.}~\bibnamefont{Marte}}, \bibnamefont{and}
  \bibinfo{author}{\bibfnamefont{P.}~\bibnamefont{Zoller}},
  \bibinfo{journal}{Phys. Rev. Lett.} \textbf{\bibinfo{volume}{72}},
  \bibinfo{pages}{997} (\bibinfo{year}{1994}).

\bibitem[{\citenamefont{Lawall and Prentiss}(1994)}]{lawall94}
\bibinfo{author}{\bibfnamefont{J.}~\bibnamefont{Lawall}} \bibnamefont{and}
  \bibinfo{author}{\bibfnamefont{M.}~\bibnamefont{Prentiss}},
  \bibinfo{journal}{Phys. Rev. Lett.} \textbf{\bibinfo{volume}{72}},
  \bibinfo{pages}{993} (\bibinfo{year}{1994}).

\bibitem[{\citenamefont{Cubel et~al.}(2005)\citenamefont{Cubel, Teo,
  Malinovsky, Guest, Reinhard, Knuffman, Berman, and Raithel}}]{cubel05}
\bibinfo{author}{\bibfnamefont{T.}~\bibnamefont{Cubel}},
  \bibinfo{author}{\bibfnamefont{B.~K.} \bibnamefont{Teo}},
  \bibinfo{author}{\bibfnamefont{V.~S.} \bibnamefont{Malinovsky}},
  \bibinfo{author}{\bibfnamefont{J.~R.} \bibnamefont{Guest}},
  \bibinfo{author}{\bibfnamefont{A.}~\bibnamefont{Reinhard}},
  \bibinfo{author}{\bibfnamefont{B.}~\bibnamefont{Knuffman}},
  \bibinfo{author}{\bibfnamefont{P.~R.} \bibnamefont{Berman}},
  \bibnamefont{and} \bibinfo{author}{\bibfnamefont{G.}~\bibnamefont{Raithel}},
  \bibinfo{journal}{Phys. Rev. A} \textbf{\bibinfo{volume}{72}},
  \bibinfo{pages}{023405} (\bibinfo{year}{2005}).

\bibitem[{\citenamefont{S{\o}rensen et~al.}(2006)\citenamefont{S{\o}rensen,
  M{\o}ller, Iversen, Thomsen, Jensen, Staanum, Voigt, and
  Drewsen}}]{sorensen06}
\bibinfo{author}{\bibfnamefont{J.~L.} \bibnamefont{S{\o}rensen}},
  \bibinfo{author}{\bibfnamefont{D.}~\bibnamefont{M{\o}ller}},
  \bibinfo{author}{\bibfnamefont{T.}~\bibnamefont{Iversen}},
  \bibinfo{author}{\bibfnamefont{J.~B.} \bibnamefont{Thomsen}},
  \bibinfo{author}{\bibfnamefont{F.}~\bibnamefont{Jensen}},
  \bibinfo{author}{\bibfnamefont{P.}~\bibnamefont{Staanum}},
  \bibinfo{author}{\bibfnamefont{D.}~\bibnamefont{Voigt}}, \bibnamefont{and}
  \bibinfo{author}{\bibfnamefont{M.}~\bibnamefont{Drewsen}},
  \bibinfo{journal}{New J. Phys.} \textbf{\bibinfo{volume}{8}},
  \bibinfo{pages}{261} (\bibinfo{year}{2006}).

\bibitem[{\citenamefont{Unanyan et~al.}(1999)\citenamefont{Unanyan, Shore, and
  Bergmann}}]{unanyan99}
\bibinfo{author}{\bibfnamefont{R.~G.} \bibnamefont{Unanyan}},
  \bibinfo{author}{\bibfnamefont{B.~W.} \bibnamefont{Shore}}, \bibnamefont{and}
  \bibinfo{author}{\bibfnamefont{K.}~\bibnamefont{Bergmann}},
  \bibinfo{journal}{Phys. Rev. A} \textbf{\bibinfo{volume}{59}},
  \bibinfo{pages}{2910} (\bibinfo{year}{1999}).

\bibitem[{\citenamefont{Kis and Renzoni}(2002)}]{kis02}
\bibinfo{author}{\bibfnamefont{Z.}~\bibnamefont{Kis}} \bibnamefont{and}
  \bibinfo{author}{\bibfnamefont{F.}~\bibnamefont{Renzoni}},
  \bibinfo{journal}{Phys. Rev. A} \textbf{\bibinfo{volume}{65}},
  \bibinfo{pages}{032318} (\bibinfo{year}{2002}).

\bibitem[{\citenamefont{Pachos and Beige}(2004)}]{pachos04}
\bibinfo{author}{\bibfnamefont{J.~K.} \bibnamefont{Pachos}} \bibnamefont{and}
  \bibinfo{author}{\bibfnamefont{A.}~\bibnamefont{Beige}},
  \bibinfo{journal}{Phys. Rev. A} \textbf{\bibinfo{volume}{69}}
  (\bibinfo{year}{2004}), \bibinfo{note}{033817}.



\bibitem{dasgupta06}
S. Dasgupta and D.A. Lidar, arXiv:quant-ph/0612201.

\bibitem[{\citenamefont{Jaksch et~al.}(2000)\citenamefont{Jaksch, Cirac,
  Zoller, Rolston, Cote, and Lukin}}]{Jaksch00}
\bibinfo{author}{\bibfnamefont{D.}~\bibnamefont{Jaksch}},
  \bibinfo{author}{\bibfnamefont{J.~I.} \bibnamefont{Cirac}},
  \bibinfo{author}{\bibfnamefont{P.}~\bibnamefont{Zoller}},
  \bibinfo{author}{\bibfnamefont{S.~L.} \bibnamefont{Rolston}},
  \bibinfo{author}{\bibfnamefont{R.}~\bibnamefont{Cote}}, \bibnamefont{and}
  \bibinfo{author}{\bibfnamefont{M.~D.} \bibnamefont{Lukin}},
  \bibinfo{journal}{Phys. Rev. Lett.} \textbf{\bibinfo{volume}{85}},
  \bibinfo{pages}{2208} (\bibinfo{year}{2000}).

\bibitem{Ohlsson02}
N. Ohlsson, R. K. Mohan and S. Kr\"{o}ll, Opt. Comm. \textbf{201},
71 (2002).

\bibitem[{\citenamefont{Bouchoule and M{\o}lmer}(2002)}]{Bouchoule02}
\bibinfo{author}{\bibfnamefont{I.}~\bibnamefont{Bouchoule}} \bibnamefont{and}
  \bibinfo{author}{\bibfnamefont{K.}~\bibnamefont{M{\o}lmer}},
  \bibinfo{journal}{Phys. Rev. A} \textbf{\bibinfo{volume}{65}},
  \bibinfo{pages}{041803} (\bibinfo{year}{2002}).
\end{thebibliography}
\end{document}